\documentstyle[multicol,pre,aps,epsfig]{revtex}

\begin{document}
\title{Block persistence}

\author{St\'ephane Cueille \and Cl\'ement Sire}
\address{Laboratoire de Physique Quantique (UMR C5626 du CNRS),\\
Universit\'e Paul Sabatier, 31062 Toulouse Cedex, France.}
\date{February, 20 1998}

\begin{abstract}
We define a block persistence probability $p_l(t)$ as the probability that
the order parameter integrated on a block of linear size $l$ has never
changed sign since the initial time in a phase ordering process at finite
temperature $T<T_c$. 
  We argue that $p_l(t)\sim l^{-z\theta_0}f(t/l^z)$ in the 
scaling limit of large blocks, where $\theta_0$ is the global
(magnetization) persistence
exponent and $f(x)$ decays with the local (single spin) exponent 
$\theta$  for large $x$.
This scaling is demonstrated at zero temperature for the 
diffusion equation and the large $n$ model, and generically it can
be used to determine easily $\theta_0$ from simulations  of coarsening
models. We also argue that $\theta_0$ and the scaling function do
 not depend on temperature, leading to {\it a definition of $\theta$ at finite
temperature}, whereas the local persistence probability decays exponentially 
due to thermal fluctuations. We also discuss  {\it conserved models} for which 
different scaling are shown to arise depending on the value of the 
autocorrelation exponent $\lambda$. We illustrate our discussion by
extensive numerical results. We also comment on the relation between this
method  and  an
alternative definition of $\theta$ at finite temperature recently introduced
by Derrida [Phys. Rev. E 55, 3705 (1997)].  
\end{abstract}

\maketitle

\begin{multicols}{2}
\narrowtext
\section{Introduction}
Phase ordering processes \cite{brayrev94} correspond to the dynamics of 
systems quenched from a disordered high temperature state to a temperature
where the equilibrium state is ordered. Dynamics proceed through coarsening
of ordered domains, and the domain linear scale $L(t)$ diverges as
$t^{1/z}$. In the coarsening regime, nontrivial spatial and temporal 
correlations
develop,  adopting a scaling form. For instance, 
the order parameter equal time correlation function   $\langle
\varphi({\bf x},t)\varphi({\bf 0},t)\rangle=f(|{\bf x}|/L(t))$. 

Universality classes depend not only on the space dimension and the 
symmetries of the order parameter, as for static critical phenomena, but
also on the conservation laws of the dynamics. Indeed, for a single static
universality class, several dynamics can be used with the only constraint 
that they must obey detailed balance. For a scalar order parameter,
non conserved  dynamics (model A)
describe a ferromagnet, while conserved dynamics (model B)
 describe demixtion or segregation in binary alloys. 
Consequently, 
the set of dynamical critical exponents such as $z$ and $\lambda$,
 defined by 
$\langle\varphi({\bf x},t')\varphi({\bf x},t)\rangle\sim
[L(t)/L(t')]^\lambda$,
for $t'\gg t$, is not related to static exponents by any hyperscaling law.

A remarkable point is that as far as the temperature $T$ of the quench is
concerned, there are only two universality classes, namely $T=T_c$ (critical
quenches) or $T<T_c$. This is assessed by numerics, renormalization group 
results or large-$n$ expansions \cite{brayrev94}. More precisely, two point
 correlations have the same scaling (up to
multiplicative constants) for any $T<T_c$. Therefore the temperature is an
irrelevant parameter  for $T<T_c$  quenches, and one may  set $T=0$ as well.

However, the situation is not as simple if one considers quantities
involving more subtle correlations. One such quantity, which has attracted 
much interest recently, is the {\it persistence probability}, defined 
as the probability that the local order parameter at a given point ${\bf x}$
has never changed sign since the initial time
\cite{derrida94a,stauffer94,marcos95}. For instance, in simulations
of the Glauber Ising model, $p(t)$ is the fraction
of spins that have never flipped since the initial time.
 At $T=0$, $p(t)$ is the
probability that a given  point has never been crossed by a domain wall. It 
usually decays with an exponent $p(t)\propto t^{-\theta}$. For  general
nonequilibrium dynamics, $p(t)$
is the probability that a  zero-mean stochastic quantity has never changed
sign since the initial time.  The analytical study of $p(t)$ is 
difficult due to the fact that it probes the whole history of the process.
Even for simple scalar diffusion with a zero-mean random initial condition,
a nontrivial algebraic decay is found \cite{majumdar96b,derrida96a}.
At $T_c$, the persistence of the global magnetization was shown to yield
a new independent critical exponent for the Ising model \cite{majumdar96c,oerding97}.

The temperature universality of the $T<T_c$ scaling of
 correlations, corresponding to
a single fixed point $T=0$ in the renormalization group,  
seems to be broken for the local persistence $p(t)$, because  at $T>0$, 
thermal fluctuations lead to an exponential decay of $p$, in contrast to 
the power law decay at $T=0$. To address this question, 
 Derrida \cite{derrida97a} recently proposed to study persistence at finite
temperature for nonconserved Ising and Potts models by comparing two 
systems $A$ and $B$ evolving with the same thermal noise from two different
initial conditions. System $A$ is initially in a completely random
configuration whereas $B$ is in its fundamental (all spins assuming the same
value). Persistence  is now defined as the probability $r(t)$ that
 $S_i^A S_i^B(t)$
has kept a constant sign since $t=0$. The underlying idea is to discard
simultaneous flips (at the same site) in both systems, because flips
in $B$ are only due to thermal fluctuations. 

The implementation 
of this simple idea by Derrida \cite{derrida97a}, and more extensive 
simulations performed by Stauffer \cite{stauffer97a}, have shown 
that  $r(t)$ decays algebraically. The observed exponent
seems to be temperature independent and equal to the $T=0$ local
persistence exponent $\theta$, for the Ising model. Therefore, universality
seems to hold with this new definition of persistence. However, the method 
cannot be used for {\it conserved models}, as system $B$ would not 
evolve. Since Kawasaki (spin-exchange) dynamics freeze at zero 
temperature, a definition of persistence at finite temperature applying
to conserved models is required. Moreover, Derrida's definition is not easy
to generalize to continuous models. A definition involving a single system 
would be more satisfactory, as we know from the study of {\it damage
spreading} that behaviors of observables 
obtained by comparison of two systems evolving with the same noise often 
depend on the Monte-Carlo algorithm used (see below).

In a recent letter \cite{cueille97a}, we 
 introduced the notion of block persistence as a very natural method to 
give a temperature independent and intrinsic definition of
 the persistence exponent. The method is in a way an {\it \`a la Kadanoff}
  implementation of the renormalization group ideas underlying the 
universality of  correlations. The block persistence probability $p_l(t)$
is the standard persistence probability for a coarse-grained variable 
obtained by integrating the order parameter on a block of linear size $l$.
In \cite{cueille97a}, we  argued that the large $l$ scaling of 
$p_l(t)$ is independent of $T$ and corresponds to the $T=0$ fixed point,
because increasing $l$ reduces the relative thermal 
fluctuations of the block variables.

In this article, we give a more detailed and general discussion of block 
persistence, which we illustrate with extensive simulations of different 
coarsening models. The structure of the paper is the following.
We start by reviewing in section \ref{sec:math} a few mathematical results 
needed to discuss persistence for physical models.

In sec. \ref{sec:comparison}, we comment further on Derrida's comparison
method and check its intrinsicality. We show that even if the persistence
exponent does not seem to depend on the algorithm used, the cross-over 
to $T_c$ and the $T>T_c$ behavior of the persistence probability is
completely different for heat bath and Glauber dynamics. 
 
In sec. \ref{sec:T0}, we start from a general discussion of global
persistence below $T_c$,  and define block persistence as a natural way to
include in a single framework the global and the local persistence exponent,
through   its scaling
for $l\to \infty$ with $l/L(t)$ fixed  
at $T=0$ and $T>0$. At $T=0$, we explicitly prove the postulated 
scaling form for the diffusion equation and the large-$n$ model.
We show that block scaling leads to an easy numerical determination
of the global persistence exponent $\theta_0$. We present numerical results
for several systems, illustrating the previous discussion.

In Sec. \ref{sec:T}, we move to finite temperature and justify  that 
the scaling should be the same as at zero
temperature, because the thermal exponential decay is eliminated in the 
scaling limit of large blocks. Thus block persistence
 provides with a definition of local 
persistence at finite temperature. We present simulations 
for the Ising and Potts models,
 illustrating temperature universality. We also discuss the $T=T_c$ case.

In Sec. \ref{sec:conserved}, we discuss the special case of conserved order
parameter dynamics. Block scaling works as for the nonconserved case, but
for an important feature: we analytically predict that 
 the scaling function should be qualitatively 
different for $\lambda=d$ and $\lambda<d$ systems.
 This prediction is confirmed by
simulations of one-dimensional models. We also present finite temperature 
simulations for the two-dimensional Kawasaki dynamics.

\section{Mathematical and general results}\label{sec:math}
Before moving to physical problems, we would like to summarize
a few useful mathematical results. 
Consider a general stochastic process $X(t)$, with $\langle X(t)\rangle=0$.
We are interested in the probability $p(t)$ that $X(t')>0$ for all 
$0\leq t <t'$. 

This is an old problem in probability theory 
\cite{slepian,blake}, but
a difficult one, and despite the large number of papers devoted to 
this subject, very few quantitative results are known, most of them concerning
 {\it stationary} and {\it Gaussian} processes, which
are completely determined by their correlator 
$C(\tau)=\langle X(t)X(t+\tau) \rangle$. With these strong restrictions, 
$p(t)$ still cannot be computed analytically, even in the large $t$ limit. 
Actually, $p[t,C(\tau)]$ is known only for very few specific correlators 
\cite{slepian,blake}. One of these correlators is $C(\tau)=e^{-a\tau}$,
which is the general correlator of a {\it Markovian} stationary Gaussian
process with the condition $C(0)=1$, for which
\begin{equation}
p(t)=\frac{2}{\pi}\arcsin(e^{-at})
\end{equation}
and $p(t)\sim (2/\pi) e^{-at}$ at large $t$.

Generally speaking, $p(t)$ and its asymptotic large $t$ decay depend
 sensitively
on the {\it whole function} $C(\tau)$ and not only on its behavior
for small or large $\tau$.  For instance,  Majumdar and Sire
\cite{majumdar96a} have considered the Gaussian process with 
$C(\tau)=(1-\varepsilon)e^{-\tau}+ \epsilon e^{-2\tau}$. 
Despite the fact that $C$ decays $\propto e^{-\tau}$ at large $\tau$ for all 
$\varepsilon<1$, simulation of the process
shows that $p(t)\propto \exp(-a(\varepsilon)t)$, where $a(\varepsilon)$ 
interpolates continuously from $1$ to $2$ when $\varepsilon$ 
is varied from $0$ to $1$.
In \cite{majumdar96a},  the Markovian correlator was used 
 as a starting point for  perturbative and variational approximations, which
 are however uncontrolled. 

The following rigorous results for any stationary Gaussian process with zero
mean are also very useful \cite{slepian,blake}: 
\begin{eqnarray}
p[t,bC(\tau)]&=&p[t, C(\tau)] \\
 p[t,C(b\tau)]&=&p[bt,C(\tau)]\\
(\forall\, \tau, \, C_1(\tau)\geq C_2(\tau))&\Rightarrow & (\forall\,t,\, p[t,C_1(\tau)]\geq p[t,C_2(\tau)])
\end{eqnarray}
 
From the first relation, we see that $p$ is completely determined by the
normalized correlator $C(\tau)/C(0)$. The second relation will be 
used to obtain scaling forms for persistence probabilities in the 
following. Finally, the third relation shows that $p(t)$ decays exponentially
in time for a  stationary Gaussian 
process with a correlator that is bracketed for all $\tau$ by two Markovian correlators
$e^{-b|\tau|}\leq C(\tau) \leq e^{-a|\tau|}$, because then $p(t)$ is  also
bracketed by two exponentials.  Most of the correlators encountered in 
physical nonequilibrium processes actually have this property in a proper 
time variable (see below). However, there might be power law prefactors
in the large $t$ decay of $p(t)$.

Of course, in nonequilibrium dynamics, 
stochastic processes are scarcely Gaussian, and, by definition, never
 stationary in physical time. However, if there is scaling relatively
to a dynamically diverging scale $L(t)$, one must have for large $t$ {\it and}
$t'$
\begin{equation}
a(t,t')=\frac{\langle X(t)\rangle X(t')\rangle}{\sqrt{\langle X^2(t)\rangle\langle X^2(t')\rangle}}  =f[L(t)/L(t')],
\end{equation}
with $f(x)=f(1/x)$. This implies  the stationarity of the process $X(t)/\sqrt{\langle
X^2(t)\rangle}$  in the 
variable $u=\ln L(t)$.

 Now if the process is Gaussian, we obtain that generically $p(u)$ decays as
$e^{-\bar{\theta} u}$ and therefore $p(t)$ decays as $L(t)^{-\bar{\theta}}$.
For most systems $L(t)\propto t^{1/z}$ and we recover the power law decay 
in time with $\bar{\theta}=z\theta$. The simplest example of such a Gaussian
process is the diffusion equation (see below). Still, because the process
is non Markovian, $\bar{\theta}$ cannot be computed analytically, and an
 independent interval approximation was 
used to predict accurately $\bar{\theta}$ \cite{majumdar96b,derrida96a}.

To end with this general discussion, we consider the following situation,
which will be of use in the study of block persistence. Consider a 
family of Gaussian processes indexed by a variable $l>0$, $\{X_l(t)\}$,
with normalized correlators $a_l(t,t')$, with the following scaling property
\begin{equation}
a_l(t,t')=h(t/l^z, t'/l^z).
\end{equation}
Then obviously $X_l(t)=X_1(t/l^z)$, leading to $p(t)=p_1(t/l^z)$.

\section{Comparison of systems}\label{sec:comparison}
Now, let us come back to coarsening processes.
Consider the nonconserved Ising dynamics. At $T=0$, $p(t)\propto
t^{-\theta}$, where $\theta$ is nontrivial and  seems to be
independent of other exponents. This is due to the
fact that spins cannot flip when they are within an ordered domain.
Flips occur only at interfaces between domains, and the slow surface tension
driven motion of these interfaces makes for the slow decay of $p$.

The situation is dramatically different at finite temperature, because
thermal fluctuations allow energetically forbidden flips. These activated 
flips occur with a decay rate  $\tau\sim e^{-\Delta E/k_BT}$, where $\Delta
E$ is a typical energy barrier to flip a spin inside a domain,  of order 
the exchange constant $J$. Therefore, these thermal flips lead
to an exponentially decaying $p(t)\propto e^{-t/\tau}$ (see
sec. \ref{sec:T}). 

Why then is there a unique scaling of correlations at finite $T<T_c$ ?
The reason is that the domain structure in the scaling regime is the same 
at any $T<T_c$. The
thermal fluctuations cancel out in the two point correlations, which
reflect only the alternation of domains of different phases. The temperature
dependence of the value of the bulk magnetization
(approximately equal to its equilibrium value), just leads to a temperature
dependent multiplicative constant in the scaling function.

From this point, it becomes clear that a simple temperature independent
definition of $\theta$ should be through the probability $r(t)$ that
{\it a given site has never changed phase}, i.e. has never been crossed
by a domain wall. At $T=0$, we clearly have $r(t)=p(t)$, and at $T<T_c$
because of the universality of the domain dynamics, $r(t)$ should have the 
same decay as at $T=0$.

 Derrida \cite{derrida97a} proposed a very clever scheme to implement this idea
for the nonconserved Ising model by simulating two systems 
$A$ and $B$ evolving with the same Monte-Carlo dynamics, with {\it the 
same thermal noise}. System $A$ is prepared in a completely random initial
condition, whereas $B$ is prepared in the fundamental state (all spins 
equal to one). Then both systems are updated simultaneously using
the heat bath algorithm with the same random number $z$  at the same site $i$:
\begin{eqnarray}
S_i^A(t+\Delta t)&=& \mbox{sign}\left[ \frac{1+\tanh (\beta \sum_i
S_i^A(t))}{2}-z\right]\\
S_i^B(t+\Delta t)&=& \mbox{sign}\left[ \frac{1+\tanh (\beta \sum_i
S_i^B(t))}{2}-z\right].
\end{eqnarray}

Then, the fraction of persistent spins $r(t)$ is defined as the fraction of
sites for which $S_i^AS_i^B$ has kept a constant sign since $t=0$.
It means that we discard flips that occur simultaneously in both systems,
because flips in system $B$ are purely thermal fluctuations, as there 
is a single $+$ phase. Accordingly, Derrida found that at finite temperature
$T<T_c$, $r(t)\propto t^{-\theta}$, with $\theta$ consistent with the $T=0$
persistence exponent. This was confirmed by extensive simulations performed
by Stauffer \cite{stauffer97a}. 

However, this practical definition of persistence for the Ising model
is not completely satisfactory. First, it cannot be directly
adapted to continuous models.  Indeed, for a continuous order parameter the probability of a simultaneous flip in both
systems will be zero in continuous time. 

A further restriction is that the method cannot be used for conserved 
dynamics, as the Kawasaki spin-exchange dynamics, because system $B$ 
would not evolve from a uniform initial condition. There is no proper 
initial condition for system $B$. This restriction is important, because
Kawasaki dynamics cannot be studied at zero temperature, and therefore 
a definition of persistence at finite temperature is required.

Finally, one would be more satisfied to get an intrinsic definition of
persistence. The comparison of two systems evolving from different initial
conditions has attracted much attention, especially in relation to the
notion of damage spreading \cite{hinrichsen97a}. It was soon realized that the behavior observed 
depends on the implementation of the Monte-Carlo algorithm. Therefore, one
 could fear that Derrida's definition may work only with the heat bath 
algorithm. To check this, we performed simulations with the heat bath
algorithm and the Glauber algorithm. We find that the $T<T_c$ behavior 
is the same for both dynamics. However, quite interestingly, the $T\geq T_c$
behavior of $r(t)$ is completely different.

For heat bath dynamics, for $T>T_c$, $r(t)$ reaches a plateau. This
 corresponds to the fact observed by Derrida and Weisbuch \cite{derrida87}
that above $T_c$ two systems evolving with this algorithm become identical 
within a finite time. When  $T\to T_c$, this plateau crosses over to a power
 law $r(t)\sim t^{-\theta_c}$. From simulations at $T_c$ we find $\theta_c
 \approx 0.9$, but $\theta_c$ can also be extracted from a scaling analysis 
of the cross-over for $T\to T_c^{+}$.  At finite $T>T_c$, there are no
 domain walls. Starting from an infinite temperature state with a
 correlation length $\xi=0$, 
$\xi$ increases to reach its equilibrium value $\xi_{eq}$.
In the vicinity of $T_c$, $\xi_{eq}\sim (T-T_c)^{-\nu}$  is very large. 
Therefore, at early times, for $\xi(t)\ll (T-T_c)^{-\nu}$ the system behaves 
as if it were to reach a critical (infinite $\xi$) equilibrium state, 
i.e. as if it were at $T_c$, and $\xi(t)\sim t^{1/z_c}$ while 
$r(t)\sim t^{-\theta_c}$. Deviations from this
power law behavior appear only at late times when $\xi(t)$ approaches the
 finite value $\xi_{eq}$ and $r(t)$ reaches a plateau. Consequently we expect the scaling form  
\begin{equation}
r(t)\sim (\xi_{eq})^\alpha g[\xi(t)/\xi_{eq}]\sim t^{-\theta_c} f[t(T-Tc)^{\nu z_c}],
\end{equation}
where $f(x)\propto x^{\theta_c}$ when $x\to \infty$ and $f(x)$ tends to a
constant  when $x\to 0$. 
\begin{figure}
\begin{center}
\epsfig{figure=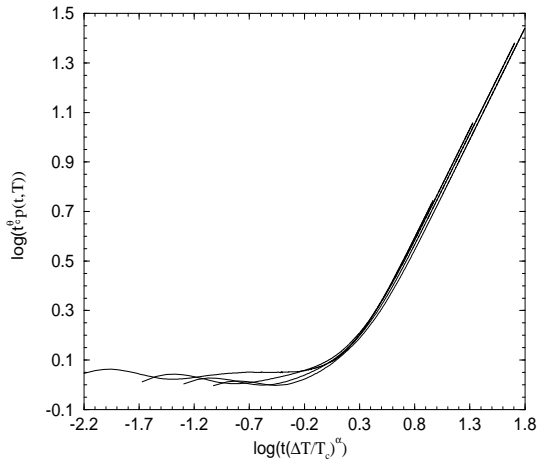, width=0.9\linewidth}
\caption{Scaling behavior of the fraction of persistence spins $r(t)$, from 
Derrida's definition, 
for the Ising model with {\it heat bath dynamics} when $T\to T_c$. Simulations
were carried out on a $1000^2$ lattice and 20 samples were averaged. 
The data collapse is obtained with $\theta_c=0.9$,
$\alpha=\nu z_c$, with the exact value $\nu=1$, and $z_c=2.17$. The scaling
function goes to a constant at small argument and diverges as a power law 
at large argument.}
\label{fig:platscal}
\end{center}
\end{figure}

\begin{figure}
\begin{center}
\epsfig{figure=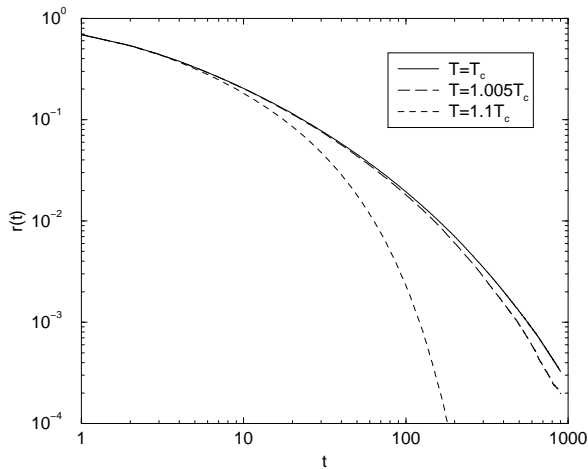, width=0.9\linewidth}
\caption{Decay of the fraction of persistent spins $r(t)$, with Derrida's
definition, for the Ising model with Glauber dynamics at $T=T_c$, $1.005T_c$
and $1.1T_c$, from simulations on a $1500^2$ lattice (20 samples).
 Above $T_c$, $r(t)$ decays faster than any power law, in contrast with the 
heat bath dynamics. At $T_c$, $r(t)$ also seems to decay faster than
algebraically, but one cannot positively rule out a power law decay with a
large exponent $\theta_c$ bigger than $2.4$ (the extrapolated exponent from
our data). }
\label{fig:glauber}
\end{center}
\end{figure}

This scaling behavior is illustrated in figure \ref{fig:platscal},  which
shows results of simulations of heat bath dynamics for the two
dimensional Ising model at different temperatures above $T_c$. The best scaling
is obtained  with $\theta_c=0.9$ ($\nu=1$ and $z_c=2.17$). It is quite 
surprising to obtain a new 
exponent at $T_c$ (see the discussion for block persistence below), and one 
should wonder whether this exponent is universal or if it depends on the 
chosen algorithm.  

If we consider another frequently used algorithm, the Glauber dynamics
which corresponds to the update rule,
\begin{equation}
S_i(t)= S_i(t)\times\mbox{sign}\left[\frac{1+\tanh (\beta S_i\sum_j
S_j(t))}{2}-z\right] 
\end{equation}
and using the same $z$ at the same site for system $A$ and $B$, 
we find a very
different behavior of $r(t)$ above $T_c$. As shown on figure
\ref{fig:glauber}, it now decays faster than any
power law. At $T_c$, it is difficult to distinguish from our simulations 
performed on a $1500^2$ lattice, averaging over 20 samples, whether $r(t)$
decays exponentially or as a power law with an exponent $\theta_c$ bigger 
than $2.4$ (the value extrapolated from our data).  
Anyway, we do not find the exponent  $\theta_c=0.9$ 
found for heat bath dynamics. Thus, this exponent is not intrinsic, neither 
is the $T>T_c$ behavior of $r(t)$ (similar results have also been found by 
Hinrichsen and Antoni \cite{hinrichsen98a}).

This illustrates the kind of problems that can be encountered using
observables defined by the comparison of two systems. Note however that,
 as said before, the large $t$ decay of $r(t)$ below $T_c$ is the same
for both dynamics, and therefore seems to be intrinsic. The cross-over in
the vicinity of $T_c^-$ will be, of course, different. One could be tempted
to relate the different behavior obtained above $T_c$ to damage spreading
properties of the dynamics. The question of damage spreading is to know whether
the distance (in configuration space) of two copies of a same system,
evolving from two slightly different initial conditions,  diverges (damage is
said to spread), or keeps bounded (damage is said to heal). For the $2d$ Ising model, damage heals 
for heat bath dynamics and spreads for Glauber dynamics. However, 
in $1d$, damage heals for both dynamics, and we have checked that both
dynamics lead to an exponential decay of $r(t)$. Therefore, the absence of 
damage spreading is not a sufficient condition to obtain the saturation of
$r(t)$, which seems to be due to a very specific property of heat bath
dynamics in $2d$. 

In the rest of the article, we describe a completely 
different approach to finite
temperature persistence, involving a {\it single system}, 
which therefore avoids
such difficulties and can be applied to continuous and conserved models.

\section{Block persistence at zero temperature}\label{sec:T0}
We consider  the nonequilibrium dynamics of  a {\it nonconserved}  
order parameter $\varphi({\bf x},t)$, which can be either continuous 
or discrete. The case of a locally conserved order parameter will be
postponed to section \ref{sec:conserved}. We first discuss block 
persistence at $T=0$, because as we will see later the scaling properties
of $p_l$ are temperature independent.

 Actually, introducing block
persistence is also quite natural at zero temperature, because it provides 
a link between the global and the local persistence probabilities. Block
 scaling will be shown below to be a very effective way of determining 
the global persistence exponent $\theta_0$, corresponding
to  the global order parameter $M(t)=\int \varphi({\bf x},t)d{\bf x}$.
This global persistence exponent has been extensively studied at $T_c$ 
\cite{majumdar96c,oerding97,stauffer96a,schuelke97a}. Here, we would like to thoroughly discuss
the $T<T_c$ case. We shall assume $T=0$, but the discussion would be the 
same at any $T<T_c$. 

\subsection{Global persistence}\label{sec:global}   
First, we remark that while  the stochastic process 
$\{\varphi({\bf x},t)\}$ (at a given point) is
generally speaking both non Gaussian and non Markovian, $\{M(t)\}$ is always 
Gaussian in the thermodynamic limit. Indeed, the vector 
$(M(t_1),..,M(t_n))$ is the sum of an infinite number 
of variables $(\varphi({\bf x},t_1)$,..., $\varphi({\bf x},t_n))$. Since the 
correlation length for $\varphi$ is finite at $t_1,...,t_n$,
the $\varphi$ vectors have short range correlations and
 the central limit theorem 
entails that the magnetization vector is Gaussian,
 for every choice of an arbitrary 
number $n$ of times. This proves that the whole {\it process} $\{M(t)\}$ is
Gaussian (which is stronger than just saying that $M(t)$ is Gaussian at any 
$t$). For a finite system, there are non Gaussian corrections due to the fact 
that the number of independent contributions to the magnetization vector is 
finite and of order $V/L(t_n)^d$, if  $t_n$ is the largest of the $t_i$.

Therefore, for an infinite system, the global persistence  
probability is completely determined by the two-times correlator 
\begin{equation}
a_g(t,t')=\langle M(t) M(t')\rangle/\sqrt{\langle M^2(t)\rangle\langle
M^2(t')\rangle}
\end{equation}
The analytical determination
of $\theta_0$ is consequently simpler in principle than for $\theta$ 
(for $\theta$, a nonlinear Gaussian approximation was used by Majumdar and Sire
\cite{majumdar96a}). At $T=T_c$, Majumdar {\it et al}
\cite{majumdar96c}  have been 
able to compute an $\varepsilon$ expansion of the the global exponent 
$\theta_c$ for model $A$.  For $T<T_c$, there is no  
natural perturbation parameter as $\varepsilon$. The analytical study 
of $\theta_0$ can be performed using the methods of \cite{majumdar96a} in
dimensions $d\geq 3$ \cite{sire96un}.  

Interestingly there is a 
relation between the autocorrelation exponent $\lambda$ and  $\theta_0$ 
{\it when $M$ is a Markov process} 
\begin{equation}\label{eq:t0_scal_law}
\theta_0 z =\lambda - d/2.
\end{equation}
This relation is the consequence of the scaling of correlations 
and  is the counterpart of a similar scaling law at $T_c$ 
\cite{majumdar96c}.

To show Eq. (\ref{eq:t0_scal_law}), we use the fact that the Gaussian
process $M(t)/\sqrt{M(t)}$ is stationary in the scaling limit as a
 function of $u=\ln L(t)$ (see sec. \ref{sec:math}), with
\begin{equation}
a_g(t,t')=f[L/L']=c(|u-u'|).
\end{equation}
The two point correlator $C({\bf k},t,t')=L^\lambda L'^{d-\lambda} 
g(k L')$ in the scaling regime for $t'\gg t$. If $g(0)=O(1)$, which is
the case for nonconserved models, $\langle M(t) M(t')\rangle \propto 
[L/L']^\lambda L'^d$ while $\langle M^2(t) \rangle \propto L^d$, and
 we obtain,
\begin{equation} \label{eq:ag_asympt}
a_g(t,t')\sim \left[\frac{L}{L'}\right]^{\lambda-d/2} \sim
\left[\frac{t}{t'}\right]^{(\lambda-d/2)/z},
\mbox{ for } t'\gg t
\end{equation}

Up to this point, the results are general and valid for a non Markovian
process. Now, if  the normalized $M$ is Markovian, then necessarily
 $c(|u-u'|)=\exp (-\theta_0z
|u-u'|)$ (see sec. \ref{sec:math}). In other words $f(x)=x^{-\theta_0z}$ 
for all $x\geq 1$, and since Eq.  (\ref{eq:ag_asympt}) 
expresses that $f(x) \sim
x^{(d/2-\lambda)}$ for $x\gg 1$,  we obtain Eq. (\ref{eq:t0_scal_law}).
Note that  the lower bound 
$\lambda\geq d/2$ proposed by Fisher and Huse \cite{fisher88} ensures that
$\theta_0$ is nonnegative. 
 Below, we shall demonstrate that $M$ is Markovian for
the  $T=0$ one dimensional Glauber model, and we shall find that
$\theta_0z=1/2=\lambda-d/2$ ($\lambda=1$), but, generally speaking,
 Eq. (\ref{eq:t0_scal_law}) is violated because $M(t)$ is non Markovian 
and $f(x)$ is not a pure power
law. The Markovian value of  $\theta_0$ is neither an  upper nor
a lower bound (see  numerical results in section 
\ref{sec:glob_res}).  The only general bound we have is 
\begin{equation}
\theta_0 \leq \theta,
\end{equation}
since obviously to flip the magnetization one has to flip single spins.

The direct determination of $\theta_0$ is  quite difficult.
 One has to record the time when the magnetization first changes 
sign, for a very large number of runs, which  limits drastically 
the sample size. Cornell and Sire \cite{cornell97} simulated the two
dimensional Ising model on a $L=8$ to $128$ lattice, and were obliged to use a
finite-size scaling analysis that did not prove very conclusive, leading to 
a large uncertainty on the  value of $\theta_0\approx 0.06\sim 0.11$.
We shall see below         
that block persistence, which we now define, 
 leads to a much easier determination of $\theta_0$.

\subsection{Block persistence}
The idea is to define  a
more general quantity, the {\it block persistence probability}, that
coincides with the global and the local persistence in different limits.
The procedure is very natural: we consider a coarse-grained variable 
$\varphi_l({\bf x},t)$, obtained by integrating   scales smaller than $l$.
The simplest procedure is to integrate $\varphi$ over {\it a block of linear
 size $l$},
as will be done for numerical simulations of lattice spin models.
Alternatively, one can also eliminate Fourier modes of wavelength smaller
 than $l$, as will be more convenient for the 
analytical treatment of continuous
models. The block persistence probability $p_l(t)$ is just the persistence
probability for the coarse-grained variable.
For $l=\infty$ we recover the global persistence, while for $l=0$ (or $1$ on
a lattice), we get the local persistence. 

Now, for finite $l$, the time dependence of $p_l$  interpolates between the two
exponents $\theta$ and $\theta_0$. Indeed, at early times, when
$L(t)\ll l$, the system effectively sees  infinite blocks, and $p_l(t) \propto 
t^{-\theta_0}$. 
Then for $L(t)\gg l$, blocks behave as single spins, and 
$p_l(t)\sim c_l t^{-\theta}$. Therefore, we expect a scaling form of
$p_l(t)$ for $l\to \infty$ with a fixed ratio $l/L(t)$
\begin{equation}\label{eq:block_scal}
p_l(t)\sim l^{-\alpha}g(L(t)/l)=l^{-\alpha}f(t/l^z),
\end{equation}
where $f(x)\propto x^{-\theta_0}$ when $x\to 0$ and 
$f(x)\propto x^{-\theta}$ when $x\to \infty$.   $\alpha$ must be equal
to $z\theta_0$ because  for finite $t$ and $l\to \infty$, $p_l(t)$ must tend
to  the global persistence probability.

This scaling form can  be demonstrated  for two analytically
tractable models closely related, namely the diffusion equation and 
the large-$n$ limit of the $O(n)$ nonconserved model. The reason is that 
in both models, the process $\{\varphi({\bf x},t)\}$ is Gaussian,
 entailing that all coarse-grained variables are also Gaussian, and
 $p_l(t)$  depends solely on the normalized 
correlator
\begin{equation}
 a_l(t,t')=\frac{\langle \varphi_l({\bf x},t) \varphi_l({\bf x},t')\rangle}{
\sqrt{\langle\varphi_l^2({\bf x},t)\rangle \langle \varphi_l^2({\bf x},t')}
\rangle},
\end{equation}
 which can be computed analytically.

\subsection{Diffusion equation}
The diffusion equation may be the simplest example of nonequilibrium
dynamics. It is not really a coarsening model, because of the absence of 
domain walls, due to the linearity of the equation.
Consider a scalar field $\varphi$ evolving according to 
\begin{equation}
\frac{\partial \varphi}{\partial t}=\nabla^2 \varphi, 
\end{equation}
from a random initial condition with zero mean $\langle \varphi\rangle=0$ and 
short range correlations 
$\langle\varphi(\vec{x},0)\varphi(\vec{x}',0)\rangle=
\Delta \delta(\vec{x}-\vec{x}')$. For this model, the global magnetization
is conserved, leading to $\theta_0=0$.

Integrating the equation in Fourier space, we obtain the Fourier transform
of the correlator
\begin{equation}
C({\bf k},t,t')=\langle\tilde{\varphi}({\bf k},t)\tilde{\varphi}(-{\bf k},t)\rangle=
\Delta e^{-k^2(t+t')}.
\end{equation}
Computing the two time correlator $C(t,t')=\sum_{{\bf k}} C({\bf k},t,t')$
leads to the normalized correlator 
\begin{equation}
a(t,t')=\frac{C(t,t')}{\sqrt{C(t,t)C(t',t')}}=\left(\frac{4tt'}{(t+t')^2}\right)^\frac{d}{4}.
\end{equation}
This correlator yields a nontrivial persistence exponent, which can be
reproduced with an excellent precision using an independent interval 
approximation (IIA) \cite{majumdar96b,derrida96a}

As remarked before, considering blocks of size $l$ is
 equivalent to introduce an upper cut-off in Fourier space 
$\lambda\sim 1/l$, and to consider $\varphi_l = (1/\sqrt{V})
\sum_{|{\bf k}|<\lambda} \tilde{\varphi}({\bf k}) \exp(i{\bf k}.{\bf x})$.
 The correlator of the corresponding block 
variables is,
\begin{equation}
C_l(t,t')=\langle \varphi_l({\bf x},t) \varphi_l({\bf x},t')
\rangle=\sum_{|{\bf k}|<\lambda} C({\bf k},t,t')
\end{equation} 

The $\varphi_l$ variables are Gaussian, and the behavior of $p_l(t)$ 
depends only on  the normalized correlator $a_l(t,t')$.
\begin{equation}
a_\lambda(t,t')=\left(\frac{\sqrt{tt'}}{t+t'}\right)^\frac{d}{2}
\frac{F(\lambda^2 (t+t'))}{\sqrt{F(\lambda^2 t)F(\lambda^2 t')}}
=H(\lambda^2t,\lambda^2 t')
\end{equation}
with $F(x)= \int_0^x y^{d-1}e^{-y^2}dy$. From the final discussion of 
sec. \ref{sec:math}, we have 
$p_l(t)=p_1(t/l^2)$, which is precisely the scaling form assumed from
physical arguments, with $\alpha=0$.

The probability  $p_l(t)$ cannot be explicited, but we can obtain 
its asymptotic behavior.
For $t,t'\gg l^2$, or for $\lambda\to \infty$ one has 
\begin{equation}
a_\lambda(t,t')\sim a(t,t')
\end{equation}
i.e. we recover the one point two-time normalized correlator, leading to 
a nontrivial exponent $\theta$.
 Then, in the opposite limit of large blocks
(or small times) $t,t'\ll l^2$, one has 
$a_\lambda(t,t')=1+O(\lambda^2(t+t'))$, which corresponds to a non-evolving 
field. The scaling of the block persistence probability is therefore conform 
to the general discussion above.

\subsection{Large $n$ limit}
Now, let us investigate the  $O(n)$ model in the large $n$ limit.
As usual, we start from a $n$-components vectorial order parameter 
$\vec{\varphi}$ with the Time Dependent Ginzburg-Landau dynamics,
\begin{equation}
\partial_t\varphi_\alpha= \nabla^2 \varphi_\alpha-r\varphi_\alpha -\frac{g}{n}
\varphi_\alpha \vec{\varphi}^2
\end{equation}
In the large $n$ limit, $\vec{\varphi^2}/n$ can be replaced by the average
 $\langle\varphi^2\rangle$,
 where $\varphi$ is now any component of the field, and one 
obtains a linear self-consistent equation, which reads in Fourier space,
\begin{equation}
\partial_t \tilde{\varphi}(\vec{k},t)=-(k^2+R(t)) \tilde{\varphi}(\vec{k},t), 
\end{equation}
with $R(t)=r+g\langle \varphi^2 \rangle $. 

Hence,
$\tilde{\varphi}({\bf k},t)=\tilde{\varphi}(\vec{k},0)h(t)^{-1/2}\exp (-k^2t)$ with 
$h(t)=\exp (2\int_0^t R(t')dt')$.  The self-consistence condition, 
\begin{equation}
\sum_{{\bf k}} |\tilde{\varphi}({\bf k},t)|^2= VS(t),					
\end{equation}  
with the definition 
$\varphi(\vec{k}) =(1/\sqrt{V})\int \varphi(\vec{x})d^d{\bf x}$, leads to 
the deterministic differential equation for $h(t)$,
\begin{equation}\label{eq:forh}
\frac{1}{2} \dot{h}=rh +\frac{g\Delta}{V} \sum_{{\bf k}} e^{-2k^2t}.
\end{equation}

The global magnetization $m(t)=\tilde{\varphi}(\vec{0},t)/\sqrt{V}$, 
is just given by $m(t)=m(0)/\sqrt{h(t)}$. Therefore, $m(t)$ is deterministic
 (apart from the randomness of $m(0)$), and  never changes sign, which
yields $\theta_0=0$. 

Equation (\ref{eq:forh}) can be solved using Laplace transform, but 
 we do not need to know $h(t)$ here. The two times correlator is
\begin{equation}
C({\bf k},t,t')=\langle\tilde{\varphi}({\bf k},t)\tilde{\varphi}(-{\bf k},t)\rangle=
\frac{\Delta e^{-k^2(t+t')}}{\sqrt{h(t)h(t')}}.
\end{equation} 
The Gaussian process $\bar{\varphi}=\varphi/h(t)$, has the same correlator as the 
diffusion equation. More precisely, $\bar{\varphi}$ obeys the diffusion
equation.
 Hence the rest of the demonstration is the same
as above. The persistence exponents $\theta$, $\theta_0=0$ and the 
scaling function $f$ are the same as for the diffusion equation.

In these two soluble models,  the scaling law of 
Eq. (\ref{eq:block_scal}) is valid for any $t$ and any $l$, and not only 
asymptotically in the large $l$ large $t$ limit  as will be the case 
in general. Remark also that the $\theta_0=0$ result 
recovers two different behaviors of the global 
magnetization. For the diffusion equation, 
the magnetization is exactly conserved, whereas for the large $n$ model
it relaxes deterministically to zero.

\subsection{Results for global persistence}\label{sec:glob_res}
Thanks to the scaling form of Eq. (\ref{eq:block_scal}),  it is possible to use
block scaling to compute $\theta_0$  numerically. One evaluates $p_l(t)$
for different $l$, which can be done on a single run, and then adjusts
$\theta_0$ to obtain the best data collapse.
We present here some numerical results for three different models,
illustrating the three possible cases: $\theta_0z$ equal to, bigger than or
smaller than $\lambda-d/2$. We also give a direct derivation of the exact
result $\theta=1/4$ for the one dimensional Glauber dynamics (Majumdar 
{\it et al} \cite{majumdar96c} used an interface representation of the
dynamics). Finally we show a surprising relation between $\theta_0$ 
for the one-dimensional $XY$ model with power law initial spatial correlations 
and $\theta$ for the diffusion equation.

\vspace{0.2cm}

\noindent {\it One-dimensional Glauber model \--} $1D$ coarsening is quite special, since the critical temperature is zero.
For the Glau\-ber Ising model, which is exactly soluble,  the exact
computation of $\theta$ was really difficult, while $\theta_0$ is trivial.
 since the global magnetization $M(t)$ is Gaussian at any time, and
Markovian in the
scaling limit. To 
show it, we just have to write the evolution equation for the two point 
correlation \cite{bray90a}, for $t>t'$,
\begin{equation}
2\frac{\partial C}{\partial t}(r,t,t')= C(r+1,t,t')+C(r-1,t,t')-2C(r,t,t'),
\end{equation}
with $C(r,t,t')=\langle S_r(t') S_0(t)\rangle$. Summing over $r$, we get,
\begin{equation}
2\frac{\partial \langle M(t)M(t')\rangle}{\partial t}=0,
\end{equation}
hence, $\langle M(t) M(t')\rangle=\langle M^2(\mbox{min}(t,t'))\rangle$.
Then, in the scaling regime $\langle M^2(t)\rangle\propto t^{1/2}$,
 leading to
\begin{equation}
a_g(t,t')=\left(\frac{t'}{t}\right)^{1/4}, 
\mbox{ for } t>t'
\end{equation} 
The normalized  correlator of the global magnetization is equal to 
$\exp [(u'-u)/4]$ in the variable $u=\ln t$. This proves that the Gaussian
process $M(u)$ is stationary and Markovian 
and that $p(u)\sim \exp(-|u|/4)$. In the $t$ variable we get 
$\theta_0=1/4=(\lambda-d/2)/z$, since $\lambda=1$ and $z=2$.
Remark that while $M(t)$ is Markovian, $S(t)$ at a given point is not,
neither is it Gaussian, and the computation of $\theta$ was 
a real tour de force \cite{derrida95a} which cannot be extended to other 
systems.

To check the scaling assumption of Eq. (\ref{eq:block_scal}) with the exact value of $\theta_0$, we have simulated the Glauber Ising model
on a $200000$ spins chain with block size 1, 21, 41, 61, 91. Ten samples
were averaged to obtain the final data. The data collapse with
$\theta_0=1/4$ is very good and  the scaling function has the expected 
behavior: a power law divergence with exponent $\theta_0$ at small argument
and an algebraic decay with exponent $\theta$ at large argument.

\begin{figure}
\begin{center}
\epsfig{figure=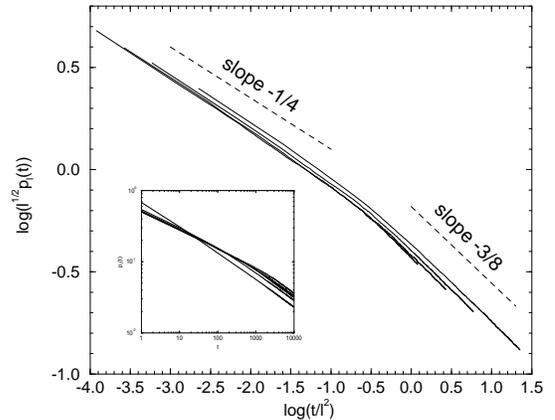, width=0.9\linewidth}
\caption{Scaling of the block persistence probability $p_l(t)$ for a one-dimensional spin chain (200000 spins, 10
samples), with block size $l=1,21,41,61,91$ (from bottom to top in the
right part of the
insert). $l=1$ is omitted in the scaling, and the data collapse improves as
the block size increases. As expected $p_l(t) \sim t^{-\theta_0}$ for 
$t\ll l^2$ and $p_l(t)\sim t^{-\theta}$ for $t\gg l^2$ (with $\theta_0=1/4$
and $\theta=3/8$). } 
\label{1d}
\end{center}
\end{figure}

\vspace{0.2cm}
\noindent {\it One-dimensional model $A$ \--} In one dimension, deterministic
and stochastic models are known to lead to different growth laws and
correlations \cite{majumdar95a}. For instance, in the one dimensional 
noiseless model $A$, domain walls have a weak attractive interaction 
decreasing exponentially with the distance, and $L(t)\propto \ln t$, whereas
in the discrete stochastic Glauber models, walls do not interact but perform
simple random walks and annihilate when they collide, leading to
$L(t)\propto \sqrt{t}$.

 Model $A$ is interesting, because it can be mapped
on a simple deterministic model of charge aggregation 
\cite{nagai86,rutenberg94,bray94a,majumdar95a,bray95a}. 
In this model,
domains of ''$+$'' and ``$-$'' phase evolve the following way.
 At each step, the smallest domain $I_0$ (length $l(I_0)$) is changed sign
 and merged with its 
two neighbors $I_1$ and $I_2$, to give a domain of length 
$l(I_0)+l(I_1)+l(I_2)$.  
To compute the domain size distribution, the sign of the
domains can be forgotten and one can easily show that no correlations
develop in the system. The mean-field equations are exact and can be solved
for the scaling function of the size distribution. In this model, 
the time variable is the minimum length $l_0$. 

Bray {\it et al.} have shown
that the local persistence exponent $\theta$ \cite{bray94a} and the 
autocorrelation $\lambda$ \cite{bray95a} have a geometrical interpretation in
this model. For instance, defining for each domain the fraction of
persistent spins $d(I)$, the new interval obtained in one step of 
the aggregation model has $d(I)=d(I_1)+d(I_2)$, and the total fraction
of persistent spin can be computed in mean field since there are no
correlations for $d$  as well.  The exact results are $\theta=0.17504588...$
and $\lambda=0.6006165...$. These exponents are solutions of implicit
nonlinear equations.

As far as the global persistence exponent is concerned, we have to consider
explicitly ``+'' and ``-'' domains, and now it is quite clear that  
correlations develop through the aggregation procedure. A naive but
instructive argument 
neglecting correlations leads to the result $\theta_0=1/2$ which is in 
contradiction with the bound $\theta_0<\theta$.

 We consider a discrete
lattice leading to integer values of $l(I)$.
To increase the minimum domain length  in the system from $l_0$
to $l_0+1$, one has to remove $n(l_0)$ domains. In each coalescence event, 
$l_0$ spins change sign leading to $\delta M=\pm 2l_0$, depending
on the sign of the domain. Now, if we assume that for small $M$, there
is no correlation between the sign and the length of domains the total
$\delta M$ is a Gaussian zero-mean variable with variance 
$\propto \sqrt{n(l_0)} 2l_0$ when $n(l_0)\gg 1$. It is easily shown 
from the mean-field equations that $n(l_0)\propto 1/l_0^2$ in the scaling
(large $l_0$ regime). Therefore, the magnetization increment is a Gaussian
variable with constant variance, and $M(l_0)$ performs a simple random 
walk, leading to $\theta_0=1/2$. This naive value  of $\theta_0$ is 
clearly wrong, being larger than $\theta$. Therefore, to obtain $\theta_0$
one should treat the correlations, which does not seem very simple.

\begin{figure}
\begin{center}
\epsfig{figure=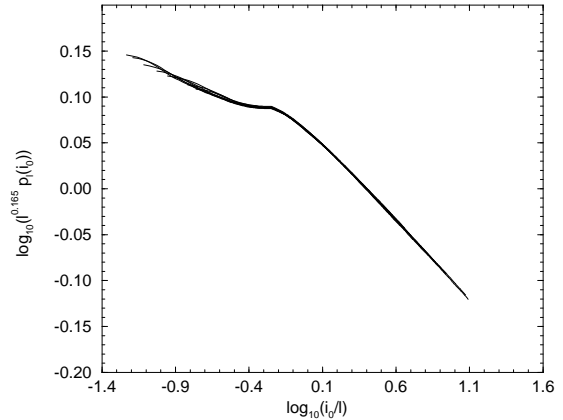, width=0.9\linewidth}
\caption{Scaling of the block  persistence probability obtained from
 simulations of the domain aggregation model equivalent to the
deterministic one-dimensional model A, on a chain with $10^6$ sites,
 $l=71,81,101,111,131,161,191,211$. Excellent scaling is found for
$\theta_0=0.165$.}
\label{fig:1d_deter}
\end{center}
\end{figure}

Numerical results for the domain aggregation model are presented in fig.
\ref{fig:1d_deter} for a $L=10^6$ chain (30 samples). The best scaling was 
obtained for $\theta_0=0.165$. The profile of the scaling function in the
cross-over region is quite different from the stochastic Glauber model.      
The Markovian scaling law would lead to $\theta_0=\lambda-d/2\approx 0.1$
($z=1$ since the time variable is the dynamical length scale). 
Hence, for this model we have $\theta_0>\lambda-d/2$. 

\begin{figure}
\begin{center}
\epsfig{figure=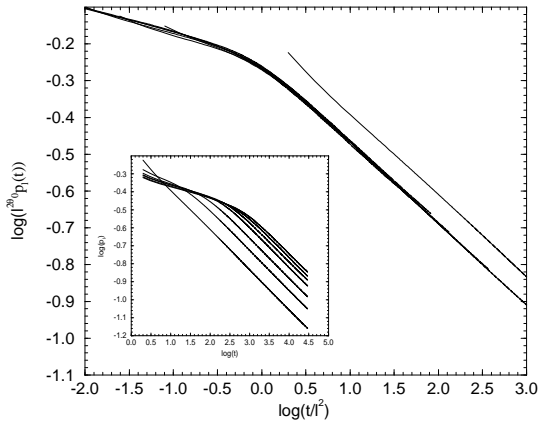, width=0.9\linewidth}
\caption{Block persistence at $T=0$ obtained from simulation of  the
nonconserved Ising model on a $2000^2$ lattice, for 
$l=1,5,9,15,19,25,$ and $31$
(from bottom to top in the insert). $p_l(t)$ decays as 
$t^{-\theta_0}$ at early
time and as  $t^{-\theta}$ at large time. Excellent scaling is then obtained
taking $\theta_0=0.09$.}
\label{zero}
\end{center}
\end{figure}

\vspace{0.2cm}
\noindent {\it Two-dimensional Glauber Ising model \--}
 $d=1$ is quite special because $T=0$ is also the critical temperature, 
 and persistence can only be defined  at
   $T=0$. Now we move to the $d=2$ Glauber  Ising model, for which block 
scaling will lead to a definition of $\theta$ at finite temperature. 
It is also interesting to determine $\theta_0$ which could only be
 roughly evaluated by the direct method \cite{cornell97} despite much
numerical effort. Comparatively, 
block scaling is a very easy and reliable method. We performed simulations 
on a $2000^2$ lattice with blocks of linear size
1,5,9,15,19,25, and 31. 20 samples were averaged to obtain the final data
presented in fig. \ref{zero}. We find excellent scaling,  
with $\theta_0=0.09$. The  uncertainty in the data collapse is roughly 
of $1\%$ on $\theta_0$. This value of $\theta_0$ is compatible with the 
range $0.06\sim 0.11$ found by Cornell and Sire \cite{cornell97}. The 
Markovian value of $z\theta_0$ would be $11/8=1.375$ ($\lambda=5/4$), and
for this model we have $\theta_0z<d-\lambda/2$.

\vspace{0.2cm}
\noindent{\it Two-dimensional Ginzburg-Landau equation \--}
We can also simulate the time dependent Ginzburg-Landau equation,
corresponding to the continuous model A,
\begin{equation}
\partial_t \varphi= \nabla^2 \varphi + a \varphi (1-\varphi^2).
\end{equation}
Starting from an uncorrelated Gaussian initial condition, one can solve
the equation using a finite differences scheme and compute $p_l(t)$
for different block sizes. Using block scaling, we can determine
$\theta_0$ and $\theta$. For both exponents, we find a value somewhat
smaller than for the Glauber Ising model: $\theta_0=0.06<0.09$ and 
$\theta=0.20<0.22$ (the value of $\theta$ has been also computed by Cornell
\cite{cornell96un}). For $\theta$, the large time decay of $p_l(t)$ shows
significant curvature and the effective exponent seems to increase with
time. The scaling function cannot be superposed with the scaling function
of the Ising model. Moreover, the fact that both 
models have different $\theta_0$  shows that the {\it two-times correlations}
of the global magnetization (which solely determine $\theta_0$) are
{\it different} in the scaling regime. This suggests that model A could be
 in a different universality class from the Ising model. This was 
also suggested by Rutenberg \cite{rutenberg96a} in a recent paper, as he 
argued that {\it model-dependent} anisotropy in the correlation function 
(due for instance to the lattice) does not vanish in the scaling regime.

\begin{figure}
\begin{center}
\epsfig{figure=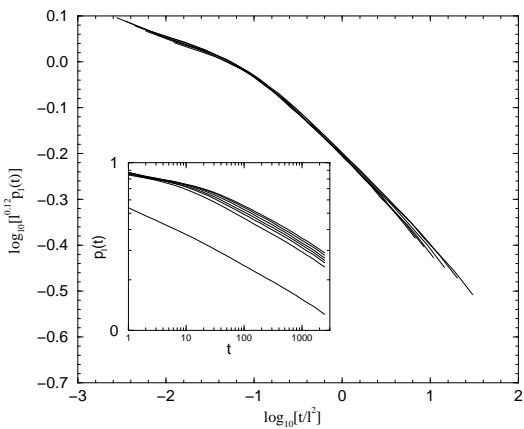,width=0.8\linewidth}
\caption{Block persistence at $T=0$ obtained from simulation of  the
time dependent Ginzburg-Landau equation at zero temperature on a $700^2$
grid, for $l=$9, 11, 13, 15, 17 and 19 (from bottom to top in the insert).
 The  scaling  presented is  obtained
taking $\theta_0=0.06$. The large time decay of $p_l(t)$ corresponds to
$\theta=0.2$, with significant curvature though.}
\end{center}
\end{figure}

\vspace{0.2cm}
\noindent{\it One-dimensional $XY$ model \--} 
The one-dimensional nonconserved $XY$ model is exactly soluble
\cite{newman90a,brayrev94}, and is quite special, as for short ranged 
initial correlation, the structure factor does not exhibit conventional 
scaling, and the growth exponent  $z=4$ in contrast with general results for 
nonconserved vector spin systems. 

The order parameter is a unitary two-dimensional vector ${\bf
\varphi}$, defined by its phase $\alpha(x,t)$.
 The equation of motion is simply a diffusion equation
\begin{equation}
\partial_t \alpha=\partial^2_{xx}\alpha,
\end{equation}
with a Gaussian initial condition, 
\begin{equation}
P(\{\alpha_k (0)\})\propto \exp \left(-\sum_k \frac{\beta_k}{2} \alpha_k(0)
\alpha_{-k}(0))\right).
\end{equation}

The correlation function $C(r,t_1,t_2)=\langle 
\cos (\alpha(r,t_1)-\alpha(r,t_2)\rangle$ depends on the initial condition. 
For a finite correlation length in the initial condition 
$\beta_k=(\xi/2)k^2$, $C(r,t,t')$ does not have the usual scaling form
\cite{brayrev94}. 
As a consequence, the normalized correlator of the magnetization is not 
a function of $t/t'$. 

Now, if the initial correlator has a power law decay $C(r,0)\sim 
r^{-\gamma/\pi}$, the general scaling form is recovered, with
\begin{equation}
C(r,t,t')=f\left (\frac{r}{\sqrt{t+t'}}\right)
\left( \frac{4t_1 t_2}{(t+t')^2}\right)^{\frac{\gamma}{4\pi}},
\end{equation}
with $f(x)\propto x^{-\gamma/\pi}$ \cite{brayrev94}. 
For $\gamma>\pi$,
the spatial correlations of the order parameter decay sufficiently fast for 
the central limit theorem to be valid (see
e.g. \cite{bouchaud90a}). Therefore, the magnetization is still Gaussian,
and the global persistence probability, now defined as the probability that
a  component of the magnetization has never changed sign, 
  is determined by the 
the normalized correlator
\begin{equation}
a_g(t,t')=\left( \frac{4t t'}{(t+t')^2}\right)^{\frac{\gamma-\pi}{4\pi}}, 
\end{equation}
which continuously depends on $\gamma$. 
This is the {\it local} correlator of   
 the diffusion equation in dimension $d=(\gamma-\pi)/\pi$. Since the
diffusing field is Gaussian, we conclude that $\theta_0$ for the one
dimensional $XY$ model with $C(r,t=0)\sim r^{-\gamma/\pi}$ is equal to 
$\theta$ for the diffusion equation in dimension $(\gamma-\pi)/\pi$, a
quite surprising result. 

\vspace{0.2cm}
Before moving to finite temperature, let us mention that in a recent 
paper \cite{hinrichsen98b}, Hinrichsen et al. have used our method to 
study persistence for a directed percolation model. Their data suggest 
that $\theta=\theta_0$. If this surprising result is confirmed, it
would be interesting to understand the exact reason for it.

\section{Block persistence at  finite temperature}\label{sec:T}
Although block persistence is useful even at zero temperature, our main 
concern remains finite temperature, which we discuss now. The main idea 
is that because the correlation length is finite at finite time, the 
relative fluctuations of the block variables vanish as $l^{-d/2}$ when
the size of the blocks is increased, and therefore the large block limit 
corresponds to no fluctuations at all, i.e. zero temperature. In fact this
picture is just a naive justification of the renormalization group flow
for coarsening. One  has to single out $T=T_c$ where the {\it relative}
fluctuations diverge, as the equilibrium magnetization vanishes, and for
which a different scaling arises.
  
\subsection{$T<T_c$}
Let us first consider $0<T<T_c$.  The difficulty in
defining a persistence exponent  comes from the fact that a spin may flip due
to thermal fluctuations, leading  to an exponential decay $p(t)\sim \exp(-
t/{\tau})$.  Indeed, at $T=0$, a spin flips only when it is crossed by an
interface between a $+$ and a $-$ domain, whereas at finite temperature, the
dominating process at late time,  when the domains are large, is the flip of a
spin within a domain  due to thermal fluctuations. Therefore, at low
temperature,  it is natural from classical kinetics intuition to expect  an
Arrhenius law $\tau\sim \exp(-\Delta {\cal E}/T)$, where $\Delta {\cal E}$ is
the energy barrier to flip a spin (or a block) within an ordered domain. As
$T\to 0$, $\tau$ diverges and $p$ crosses over to a power law.

 Arrhenius laws are common enough in physics and chemistry, and arise 
each time  a fluctuating process has to cross a finite barrier. It is
useful, though,  to work out the random process viewpoint, 
to clearly understand how $\tau$ should behave with $l$.

Let us consider a block of linear size $l$, and spin block variables
$\varphi_l$. When  $L(t)$ is large enough,   the system can be considered
locally at equilibrium inside a domain, and, since there are no long-range
correlations, 
$\langle\varphi_l(t)\rangle\approx l^d \langle \varphi \rangle_{eq}$ and 
$(\Delta \varphi_l)^2=\langle\varphi_l^2(t)\rangle-\langle
\varphi_l(t)\rangle^2\approx l^d (\Delta \varphi)^2$. Therefore the relative 
fluctuation of  $\varphi_l$ has the scaling $\Delta
\varphi_l/\langle\varphi_l\rangle \propto \sqrt{T/l^d}$.

  Thus $p_l(t)$ is
essentially the probability that the stationary  random process
$\varphi_l(t)$  with mean value of order $l^d$ and fluctuations of the same 
order has never crossed zero. In other words, it is the survival
probability of a stationary walker 
$X(t)=(\varphi_l(t)-\langle \varphi_l\rangle)/\langle\varphi_l(t)\rangle$,
 with  zero mean and a mean square fluctuation  
$\langle X^2\rangle=aT/l^d$, and an absorbing boundary at $x=1$. 

To simplify, let us assume that  $X(t)$ is Gaussian and Markovian. Then one
can write  a simple Langevin equation,
\begin{equation}
\dot{X}(t)=-\gamma X(t) +\eta(t)
\end{equation}
 with a Gaussian white noise $\eta(t)$ with
$\langle\eta(t)\eta(t')\rangle=2aT/l^d$ $\delta(t-t')$. 
In  exponential time $u=e^{2\gamma t}$, the new random variable $Y(t)=2\gamma
\sqrt{u}X$ performs a simple random walk,
\begin{equation}
\dot{Y}=\xi(t),
\end{equation}
where $\xi(t)=\eta/\sqrt{u}$ is a new Gaussian white noise, 
with $\langle \xi(u)\xi(u')\rangle=4\gamma aT/l^d\delta(u-u')$.
Hence, $p_l(u)$ is the survival probability of a
simple $1d$ random walker with diffusion coefficient $D=2aT/l^d$, starting from
$x=0$ with a moving  absorbing wall  at $x(u)= \sqrt{u}$. The survival probability is just,
\begin{equation}
S(u)=\int^{\sqrt{u}}_{-\infty} P(x,u)dx,
\end{equation}
where $P(x,u)$ is the presence probability of the walker. $P(x,u)$ is the
solution of the diffusion equation with an absorbing boundary condition 
at $x=\sqrt{u}$.  When the
wall motion is much faster than the diffusion of the walker, 
 i.e. $D\ll 1$, which corresponds  to small $T$ or large $l$ 
(small fluctuations), 
$P(x,u)$ can be well approximated by  a Gaussian distribution 
with a time dependent weight $S(u)$ \cite{krapivsky},
\begin{equation}
P(x,u)= \frac{S(u)}{\sqrt{4\pi Du}}e^{-\frac{x^2}{4Du}}, 
\end{equation}
where $S(u)$ is determined by equating the  mass loss rate with the flux of 
mass through the moving wall.
At large $u$, $S(u)$ decays with a  power
law $u^{-\beta}$ and $\beta = (4\pi D)^{-1/2}
\exp(-1/4D)$. Since $p_l(t)=S(e^{2\gamma t})$, 
we recover the heuristic
Arrhenius law with,
\begin{equation}
\tau=1/(2\gamma\beta)=\sqrt{\frac{2\pi a T}{\gamma^2 l^d}}\exp[l^d/(8 a T)] 
\end{equation}
The constant $a$ is a slowly varying function of the equilibrium correlation
 length but does not depend on $l$ for large $l$.   

 The important point is that the effective temperature
entering the Arrhenius law of the spin blocks is cut by a factor $l^d$ and
that $\tau$ diverges very quickly when $l$ is increased, leading to a fast
cross-over to the $T=0$ behavior.
Admittedly, the actual stochastic process $\varphi_l(t)$ is certainly
non-Markovian. However, for $l$ much bigger than the equilibrium 
correlation length, it is nearly Gaussian from the central limit theorem.
  Moreover, its  correlator
$C(t)=\langle\varphi_l(t)\varphi_l(0)\rangle-\langle\varphi_l(0)\rangle^2$ can be bounded by two
Markovian exponential  correlators (because there is no long range correlation
in time  at equilibrium), and thus the Arrhenius law still holds with proper
constants inserted (although the power law in the prefactor may be
modified), from the discussion of sec. \ref{sec:math}.

 For $t\ll \tau$, $p_l(t)$ is expected to
behave in a similar way as for $T=0$, and  we expect
\begin{equation}
p_l(t)\sim 
l^{-z\theta_0}f(t/l^z)\exp[-t/\tau(l,T)],
\end{equation}
with two different cross-over times.
However, in the scaling limit $l\to \infty$, $\tau$ diverges much faster
than $l^z$, entailing that the exponential part does not scale. Hence, the 
scaling form of $p_l(t)$ should be Eq. (\ref{eq:block_scal}). Moreover, from
the universality of the domain wall dynamics for $T<T_c$,
 the scaling function $g$
should be the same as at zero temperature, up to an overall temperature 
dependent multiplicative factor. As for the scaling function $f$, we have to
take into account a temperature dependent multiplicative constant in $L(t)$
(see below).   

\begin{figure}
\begin{center}
\epsfig{figure=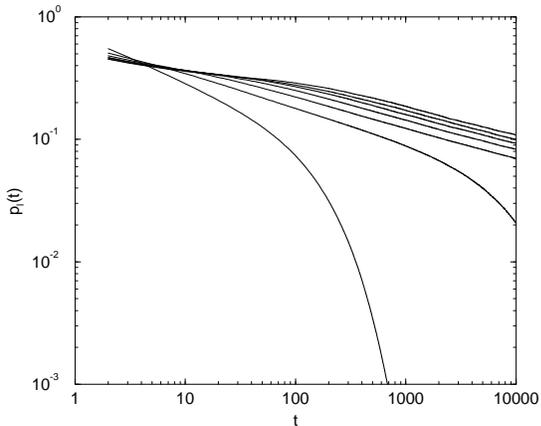, width=0.9\linewidth}
\caption{$p_l(t)$ for the two dimensional Ising model at $T=2T_c/3$, and block sizes $l=1,3,5,7,9,11,13$. 
The exponential decay of $p_l(t)\sim e^{-t/\tau}$ is clearly visible
 for $l=1$ and $l=3$, 
however for $l=5$, the exponential regime is already repelled at times
longer than the simulation time, in agreement with the expected fast
divergence $\tau\sim \exp(al^2/T)$. For $l>5$, only the power law 
zero-temperature like regime is to be seen.}
\label{fig:t2/3}
\end{center}
\end{figure}

\begin{figure}
\begin{center}
\epsfig{figure=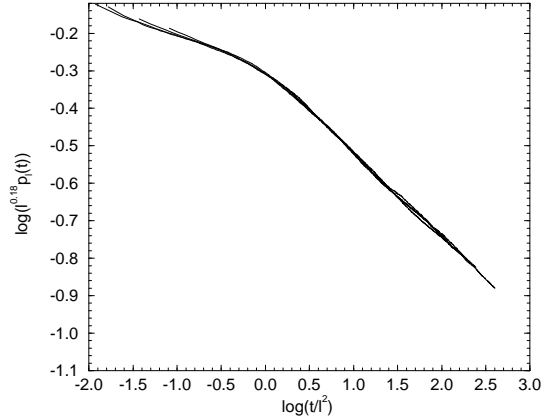, width=0.9\linewidth}
\caption{$p_l(t)$ expressed in scaling form for $T=T_c/2$, and block sizes
$l=7,9,11,13$, using the same value for $2\theta_0=0.18$ as in the $T=0$ case.
Note the similarity with the $T=0$ scaling function of fig. 1.}
\label{t0.5}
\end{center}
\end{figure}

To illustrate these ideas, we have performed simulations of the
{\it two-dimensional Glauber Ising model} at finite temperature on a $1000^2$
lattice. Figure \ref{fig:t2/3} presents results 
at $T=2T_c/3$ for blocks of size $l=1,3,5,7,9,11,13$.  The
exponential decay is clearly visible for $l=1$ and $l=3$. However, for larger
blocks, $\tau$ is bigger than the simulation time, and $p_l(t)$ has the $T=0$
behavior, with a power law decay with exponent $\theta$  fully compatible with
the $T=0$ value ($\theta=0.22$), for $t\gg l^2$, and a power law decay with
exponent $\theta_0$, for $t<l^2$, just as expected. Figure \ref{t0.5} shows
the scaling function at $T=T_c/2$ (where the approach to scaling is faster)
obtained with the zero temperature value  $\theta_0=0.09$,
  for $l=7,9,11,13$. The data collapse is really excellent. 

\begin{figure}
\begin{center}
\epsfig{figure=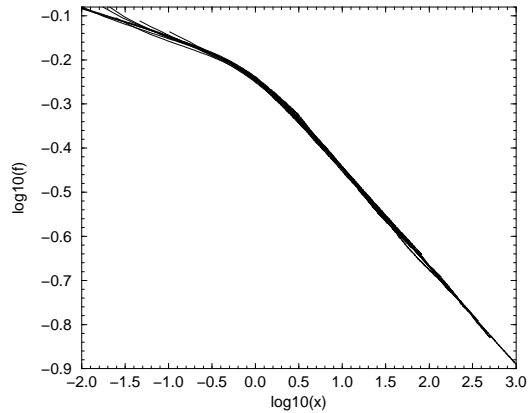,width=0.8\linewidth}
\end{center}
\caption{Universality of the scaling function for block
persistence. We show the superposition of scaling data corresponding to
$T=0$ and $T=T_c/2$. For $T=0$, we (double-log) plot 
$f=l^{2\theta_0}p_l(t)$ versus 
$x=t/l^2$, whereas  for $T=T_c/2$ we plot $f=a_1l^{2\theta_0}p_l(t)$ versus 
$x=a_2\,t/l^2$, with $a_1=1.07$ and $a_2=1.26$. The excellent superposition 
of the two scaling functions assesses the universality. }
\label{fig:univ}
\end{figure}

The temperature universality of the scaling function is illustrated in
fig. \ref{fig:univ}.  We plot the quantity $f=l^{2\theta_0}p_l(t)$
versus $x=t/l^2$ for a set of zero temperature data,
 and $f=a_1 l^{2\theta_0}p_l(t)$
versus $x=a_2t/l^2$ for  $T_c/2$ data, for blocks of size 7,9,11,13. 
The constants $a_1$ and $a_2$ are the same for all sizes, and are adjusted 
to superpose the two sets of data. $a_2$ arises from the temperature
dependence of the prefactor of $t^{1/2}$ in $L(t)$ 
(the natural time variable) and $a_1$ is the overall temperature dependent
multiplicative discussed above. The superposition obtained is really
excellent and assesses the expected universality of the scaling of block
persistence, in a very similar way as what is known for the equal-time
two-point spin correlation function \cite{brayrev94}.

Thus, block scaling leads to a 
 definition of $\theta$ at finite temperature 
 as the exponent of  the algebraic decay of
 the scaling function
$f(x)$. For the two-dimensional Ising model, the temperature independence of
 $\theta$ obtained with this method
confirms   the results obtained with Derrida's definition, but the
 universality is stronger, since the {\it whole block persistence 
scaling function } is universal. This universality arises from general 
arguments and should be observed for generic systems.  

Note that universality  would rather be expressed in terms of $L(t)$ than 
in terms of $t$.
 This is especially relevant for the {\it three dimensional Glauber Ising
model} with nearest neighbors interactions on the cubic lattice. Numerical 
simulations \cite{stauffer94,majumdar96a} at $T=0$ lead to $\theta\approx
0.17$, whereas at finite temperature our method leads to $\theta_{T>0}\approx
0.26$ in agreement with results obtained by Stauffer \cite{stauffer97a}
using Derrida's definition. For this problem, seemingly due to lattice
effects, $L(t)$ does not grow as $t^{1/2}$ at zero temperature, but as 
$t^{0.33}$ \cite{shore92}. At finite temperature, lattice effects are
overcome, and one recovers the usual growth law. Now if 
the block scaling function $g$ is universal, 
one should have the same value of $\theta z$ at any temperature. 
From numerical results we obtain $\theta z= 3.0\times 0.17=0.51$ at $T=0$
and $\theta z=0.26\times 2=0.52$ at finite $T$, which actually confirms
this {\it universality}. Note that these values of $\theta z$
are in good agreement with an approximate continuous theory \cite{majumdar96a}.

Block persistence is also very useful to study persistence for the 
{\it $q$-state Potts model}, as  zero temperature dynamics show blocking 
effects at zero temperature on the square lattice with nearest neighbor 
interactions \cite{derrida96b}. Working at finite temperature is a more
satisfactory way of overcoming blocking effects than changing the lattice
type or including next nearest neighbors interactions.
Derrida used his comparison method  to study the $q=7$ Potts model. His data
 seemed to   suggest a temperature dependence of $\theta$ \cite{derrida97a}.  

 On the basis of the present work, we would rather expect $\theta$
to be independent of $T$, at least with our definition.  To address this 
question, we have performed simulations of the $q=7$ Potts model at
$T=T_c/3$ and $T=2T_c/3$, on a $1000^2$ lattice. We have computed $p_l(t)$
for $l=$1, 3, 9, 11, 15, 19 and 25, where
 the block variables are defined through a majority rule. For both
temperature, acceptable scaling is obtained for $\theta_0=0.3$, but the
scaling is not as good as for the Ising model, and would surely  be improved
by using larger block and simulating longer times. Moreover, nonscaling
transients extend over quite a long period of time for $T=2T_c/3$. The
extraction of $\theta$ from the decay of $p_9(t)$  leads to
$\theta(2T_c/3)\approx 0.485$ and $\theta(T_c/3)=0.47$. The discrepancy 
is not really significant compared to numerical uncertainties, and is much
smaller anyway than for  Derrida's data \cite{derrida97a}, who found
$\theta(2T_c/3)\approx 0.55$ and $\theta(T_c/3)\approx 0.4$. Hence,
 $\theta$ does not seem to depend on temperature. This is confirmed by the 
comparison of the two scaling functions, which can once again be superposed
through a global rescaling. 
\begin{figure}
\begin{center}
\epsfig{figure=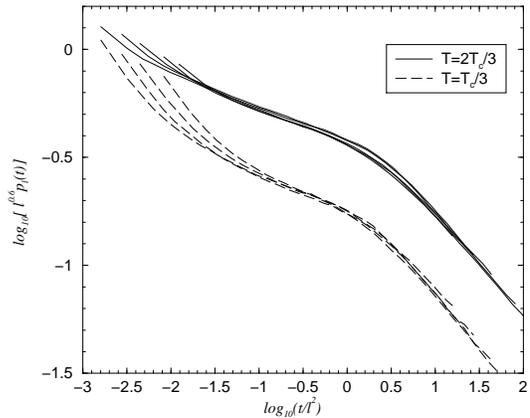,width=0.8\linewidth}
\caption{Scaling of block persistence for the $7$-state Potts model at
finite temperature $T=T_c/3$ and $T=2T_c/3$, from simulations on a $1000^2$
lattice for blocks of size $l=$9, 11, 15, 19 and 25 (13 samples). Data
are noisier than for the Ising model, but  acceptable scaling is obtained with
 $\theta_0=0.3$. }
\label{fig:potts}
\end{center}
\end{figure}

 Finally, the value of
$\theta$ compares well with zero temperature data obtained by Derrida {\it
et al.} \cite{derrida96a}. These authors simulated the next nearest neighbor 
interactions Potts model to avoid blocking effects. Data for $p(t)$ showed
significant curvature, due to the fact that the effective $z$ exponent 
increases with time, and better results were obtained for the exponent
$\varphi$ defined as $p(t)\sim L(t)^{-\varphi}\sim E(t)^\varphi$, where $E$ is the 
energy difference with the fundamental state. These authors found
$\varphi=1.01$ for the $q=7$ Potts model. Assuming  
$\varphi=\theta z$ with the asymptotic value $z=2$,
 their data lead to  $\theta \approx 0.5$,  
in acceptable agreement with our results at finite temperature.

\subsection{$T=T_c$}
The naive kinetic argument giving the scaling of the relaxation rate 
of $p_l(t)$ is bound to break down at the critical temperature for several 
reasons. We know from explicit renormalization group analysis 
\cite{hohenberg77} that $T_c$ is a fixed point for the dynamics. Hence 
in contrast with $T<T_c$, the thermal  
 decay of the persistence
probability must  scale.  Since the equilibrium magnetization is zero, one can no longer make
a distinction between a slow flip mode due to interface motion and a fast
flip mode due to thermal equilibrium fluctuations within domains. There are
no domain walls in the system, and the relevant length scale is the time
dependent correlation length $\xi(t)\sim t^{1/z_c}$. 
During the dynamics, patches of correlated spins of length $\xi(t)$ appear
in the system. These large patches have a large life time due to {\it critical
slowing down}. If we consider a block of spins inside one of these patches, 
the typical time required to flip the spin of the block should be roughly 
speaking of the order of the time required to relax a  fluctuation  of
 wave vector $\approx 2\pi/l$ in the critical equilibrium state.
 Hence, one should have $p_l(t)\sim
\exp(-\omega_c(2\pi/l)t)$ where $\omega_c(k)$ is the characteristic
critical relaxation frequency and scales as $k^{z_c}$ \cite{hohenberg77}. This leads to an
exponential decay $p_l(t) \sim \exp(-at/l^{z_c})$. Therefore, as 
predicted by the
renormalization group argument, the exponential decay of the persistence
 probability scales
and we must have
\begin{equation}
p_l(t)\sim l^{-\theta_c z_c} g(\xi(t)/l)=l^{-\theta_c z_c}f(t/l^{z_c})
\end{equation}
in the scaling limit $l\to \infty$ with a fixed ratio $\xi/l$, where 
$f(x)\sim x^{-\theta_c}$ when $x\to 0$ and $f(x) \sim e^{-ax}$ when 
$x\to \infty$. The exponent $\theta_c$ is the global persistence exponent 
at $T_c$ \cite{majumdar96c}. 

\begin{figure}
\begin{center}
\epsfig{figure=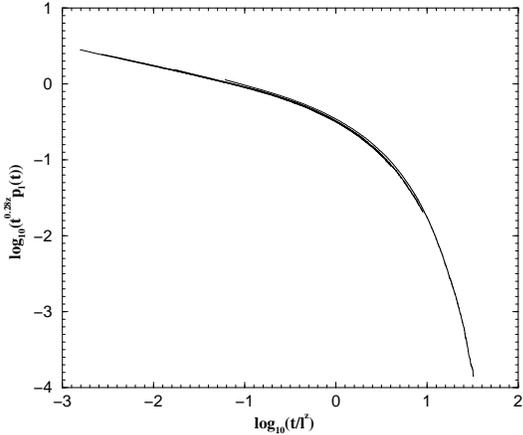,width=0.8\linewidth}
\end{center}
\caption{Scaling of the block persistence probability  at $T_c$ from
simulations of the two-dimensional Ising model on a $1000^2$ lattice 
 (15 samples) and blocks of size $l=$5, 9, 13, 17, 21 and 27. We have taken $z=z_c=2.17$ and adjusted $\theta_c$. The best
collapse is obtained for $\theta_c=0.28$. }
\label{fig:Tc}
\end{figure}

This scaling theory was checked for the two-dimensional Glauber Ising model
from simulations on a $1000^2$ lattice and blocks of size 1,5,9,13,17,21 and
27. Fifteen samples were averaged to obtain the data shown in fig. 
\ref{fig:Tc}. Excellent scaling is found with $z_c=2.17$ and
$\theta_c=0.28$, which is also in agreement with the value of $\theta_c$ 
obtained by fitting the small $x$ power law.
 The scaling function has an exponentially fast decay at large
argument, as expected. The value obtained for $\theta_c$ yields 
$\theta_c z_c \approx 0.607$, some $20\%$ bigger than the value found by 
direct determination of the global persistence probability
\cite{majumdar96c,stauffer96a,schuelke97a}, and the reason for this
discrepancy is unclear.

\section{Conserved models}\label{sec:conserved}
The amount of certitudes we have for conserved order parameter dynamics
is by far much smaller than for the nonconserved case. The large $n$ 
limit is quite pathological as it exhibits multiscaling, which is not 
observed in simulations. The only exact result is the celebrated 
Lifshitz-Slyozov-Wagner (LSW) theory \cite{lifshitz61,wagner61} for the limit of a vanishing
concentration of minority phase. In this limit, well separated droplets of
minority phase  are embedded in a matrix of majority phase. This 
spatial structure is very different from the labyrinth-like  domain
structure of the equal concentration case. The typical length scale $L(t)$
scales as $t^{1/3}$ (z=3) and one can compute the scaling function for equal
time correlations. Recently, Lee and Rutenberg \cite{lee97}, have shown 
that $\lambda=d$ for LSW. 

For finite concentrations of the minority phase, and especially for the
zero-magnetization case, the situation was more controversial. While 
the $t^{1/3}$ growth law seems well established since     the numerical work
of Huse \cite{huse86}, no conclusive result for $\lambda$ is available.
Numerical simulations of conserved models are difficult because the dynamics
are much slower than for the nonconserved models, and that corrections to 
scaling are important even at long  simulation times (see below).
 Moreover, the spin-exchange Kawasaki dynamics freeze at zero temperature, 
and simulations must be performed at finite temperature, and the standard
definition of persistence cannot be used. This explains why results for
conserved persistence can be reduced to an analytical computation of
$\theta$ in the LSW theory \cite{lee97}. In fact, in the absence of
numerical simulations, the question 
of whether the persistence probability has a power law decay or not 
is open, even if the answer is intuitively yes.

Here, using block persistence, information can be extracted from
finite temperature data, and we are able to study persistence for the
Kawasaki $d=2$ model, with the numerical limitations discussed above.
Basically, the discussion for nonconserved models can be directly 
adapted to conserved models, but for an important subtleness that
we now point out.

\subsection{Block scaling}
In the discussion of the scaling of the block persistence probability, some
additional care is required compared to the nonconserved case. 
To understand why, let us recall the discussion of Majumdar
and Huse \cite{majumdar95a} for $\lambda$.

In the scaling regime, the Fourier transform of the two points correlator 
must have the scaling form $C({\bf k},t,t')= L^{d} f(k L,L/L')$.
From the definition of $\lambda$, we have $f(x,y)\sim c y^{-\lambda}$ at large
$y$. Since the global magnetization is
constant,  $C({\bf k}=0,t,t')$ is a time-independent constant $\Delta$
 (its value depends
on the initial condition), leading to $f(0,y)=\Delta y^{-d}$ for all $y$,
 which  imposes $\lambda=d$, precisely the
 result found by Lee and Rutenberg for the LSW theory \cite{lee97}.
However, Majumdar and Huse \cite{majumdar95a} have shown numerically 
that the  one dimensional continuous $T=0$ model $B$, which can be mapped on
a deterministic domain aggregation model, has $\lambda<d$. Why does  the 
argument above fail for this model ? 

In fact, we have abusively ruled
out the possibility that $f(0,L/L')=0$. In this case, the ${\bf k}=0$ mode does
not scale and we must write more generally \cite{majumdar95a}
\begin{equation}
C({\bf k},t,t')= L^d f(k L,L/L') + f_1(k L,L/L'),
\end{equation}
 where the second term is the leading correction to scaling. If
 $f(0,L/L')=0$, this term is negligible for any finite ${\bf k}$, but not for 
${\bf k}=0$, and we have $C({\bf k}=0,t,t')=\Delta=f_1(0,L/L')$. In this case, we
 obtain no information on $\lambda$, which may be either equal to $d$ or 
 nontrivial. Conversely, if $f(0,L/L')$ is finite, then the previous 
argument  applies, and $\lambda=d$.

Coming back to block persistence, we  make the scaling
assumption of Eq. (\ref{eq:block_scal}). If we just import the result of the
nonconserved case, we find $\alpha=z\theta_0=0$, and $f(x)\to 1$ when $x\to
0$, because of the conservation law. However, this result does not always 
hold. 

The coarse-grained correlator $C_l(t,t')$ is given by,
\begin{equation}
C_l(t,t')=\int_{kl<1} \frac{d^d{\bf k}}{(2\pi)^d} C({\bf k},t,t'). 
\end{equation}
In the scaling regime, we have
\begin{equation}
C_l(t,t')= \int_{u<(l/L)} \frac{d^d{\bf u}}{(2\pi)^d}(f(u, L/L') +L^{-d} f_1(u,L/L'))
\end{equation}
Now, in the scaling limit for the persistence probability of $l\to\infty$
with fixed ratio $x=L/l$ and $y=L'/l$, we get,
\begin{equation}
C_l(t,t')= \kappa_d\int_0^{1/x} u^{d-1}f(u,x/y)du
\end{equation}
Now, if we consider the rescaled times $x$ and $y$, the small argument 
behavior is determined by the small $x$ and $y$ asymptotics of the
normalized correlator, since in this limit, corresponding to $L$
and $L'$ much smaller than $l$, the block variables are Gaussian.
If the ${\bf k}=0$ mode scales, $f(u,x/y)\to \Delta (y/x)^{d}$ when $u\to 0$,
leading to  $a_l(x,y)\to 1$.
If $f(0,x/y)=0$, we obtain,
\begin{equation}
a_l(x,y)= \left(\frac{y}{x}\right)^{d+j}\kappa(x/y),
\end{equation}
with $f(u,x/y)=u^j\kappa(x/y) + o(u^j)$.

Hence $a_l(x,y)$ is a nontrivial function of $x/y$, leading to the fact
that $p_l(t)$ scales as $c_0x^{-\theta'}$, where $\theta'$ is a nontrivial
exponent. It means that the  constant magnetization regime exists  at early 
time, but  is
confined up to a nonscaling cross-over time, and therefore does not appear
in the scaling function. Finally, the correct scaling  for $p_l(t)$, is
\begin{equation}\label{eq:nontrivl}
p_l(t) \sim l^{-z\theta'}f[L(t)/l],
\end{equation}
with $f(x)\propto x^{-z\theta}$ when $x\to \infty$, and 
$f(x)=x^{-z\theta'}$ when $x\to 0$. The exponent $\theta'$ is zero 
if the ${\bf k}=0$ mode scales, and is nontrivial otherwise. Note that 
$\theta'$ can be bigger than $\theta$ and that it is possible, although 
highly unlikely, that $\kappa(x/y)=(x/y)^{d+j}$, which would lead
to $\theta'=0$.

To check the validity of this analytical discussion, we have simulated 
two different one-dimensional conserved models, illustrating the 
two possible cases $\lambda=d$ and $\lambda<d$. 

\subsection{1D-Kawasaki dynamics}
The one dimensional spin-exchange dynamics (Kawasaki dynamics) is special,
in the sense that it does not coarsen at any temperature. Indeed, since its
critical temperature is zero, coarsening does not occur at finite
temperature, whereas the system freezes at zero temperature. However, Majumdar
{\it et al.} \cite{majumdar94a,majumdar95a} have shown that coarsening occurs  in
the $T\to 0$ limit in the rescaled time $\tau=t\exp(-4J/k_bT)$. The obtained
dynamics is equivalent to a domain diffusion model of Cornell {\it et al} 
\cite{cornell91}.  In this model, domains of length $L$ perform random walks
with a diffusion constant proportional to $1/L$ and coalesce. At small finite
temperature, this corresponds to the fact that a domain of $+$ phase 
moves through the diffusion of an isolated $-$ spin detached with
probability $\exp(-4J/k_bT)$ from a
neighboring $-$ domain and reaching the other neighboring domain after 
about $L^2$ steps of a random walk \cite{majumdar94a,cornell91}.
Majumdar {\it et al.} have argued that for this model $\lambda=d$, which
they have checked numerically \cite{majumdar94a}.

\begin{figure}
\begin{center}
\epsfig{figure=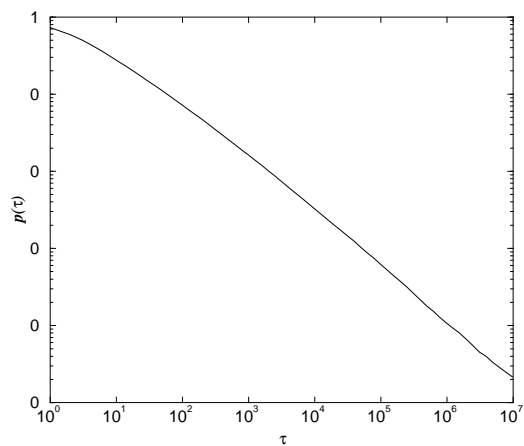,width=0.8\linewidth}
\end{center}
\caption{Numerical results for the fraction of persistent spins in the
domain model corresponding to the $T\to 0$ limit of the Kawasaki Ising
chain ($10^6$ sites, $12$ samples). We find a power law decay at large time
with $\theta=0.73$.
}
\label{fig:kawa_1D}
\end{figure}

The local persistence exponent of the one-dimensional Kawasaki dynamics 
can be defined through this
domain model. We present on fig. \ref{fig:kawa_1D} results of 
simulations of the model on a $L=10^6$ 
chain (10 samples). We observe a power law decay with
 $p(\tau)\propto \tau^{-\theta}$, with $\theta=0.73$.   To our knowledge, 
this is the first
 numerical demonstration of the existence of a
persistence exponent for a conserved model, confirming  the result obtained 
in the  LSW limit by Lee and Rutenberg \cite{lee97}. 
Note that the  persistence
exponent is  much bigger than for Glauber dynamics ($\theta=3/8$) (see below).
The complexity of the aggregation model leaves little hope of obtaining the 
exact value of $\theta$ for this model.

\begin{figure}
\begin{center}
\epsfig{figure=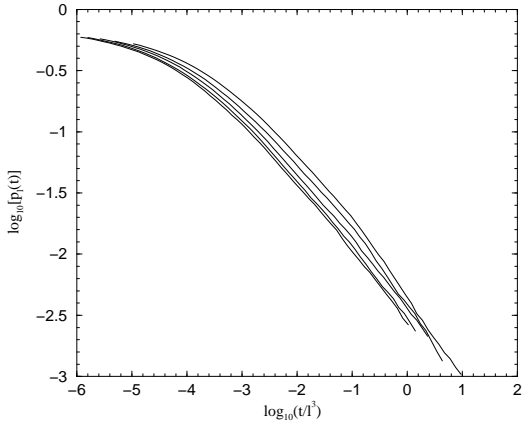,width=0.8\linewidth}
\end{center}
\caption{Block persistence probability in scaling form (with $\theta'=0$)  
for the domain model corresponding to the $T\to 0$ limit of the Kawasaki Ising
chain. Simulations were carried out for a $10^6$ sites chain (12 samples)
for blocks of size $l=$101, 131, 161, 191 and 211 (from top to bottom).
 The  approach to scaling is very slow when $l$ is increased, but it is
clearly visible that the scaling function goes to a constant ($\theta'=0$)
as expected since $\lambda=d$.
}
\label{fig:k1d_bloc}
\end{figure}

Since $\lambda=d$, we expect the naive scaling of block persistence with
$\theta'=0$. We have computed $p_l(t)$ for $l=$101, 131, 161, 191 and
211, and 
the persistence probability is presented in scaling form with $\theta'=0$ in 
fig. \ref{fig:k1d_bloc}. For increasing $l$, the  approach to scaling is
very slow and although we have used very large blocks, the data collapse is 
poor. However, it is  worse if one tries to set $\theta'\neq 0$, and one
clearly sees that $f(x)$ tends to a constant when $x\to 0$, confirming the
theoretical prediction.   

\subsection{Deterministic domain model}
Another one-dimensional conserved model, is the zero temperature
Cahn-Hilliard equation (model B). Majumdar and Huse \cite{majumdar95a} 
have shown that the dynamics could be mapped on a deterministic domain
aggregation model.  In one step of the dynamics, the shortest domain $I_0$
of length $l_0$ is localized and removed, the left (length $l_l$)
 and right (length $l_r$) neighbors are 
merged. The length  $l_0$ is dispatched between  the right ($l_{rr}$)  and left ($l_{ll}$) second neighbors 
(which have the same sign as $I_0$), according to   $l_{ll}=l_{ll}+l_{0l}$
and $l_{rr}=l_{rr}+l_{0r}$, with $l_{0r}+l_{0l}=l_0$ and
$l_{0r}:l_{0l}=l_l:l_r$. These domain dynamics reflect the fact that the
shortest domain shrinks due to  
diffusion fluxes from $I_0$ to its second neighbors through its first 
neighbors. The fluxes are proportional to $e^{-2l_0}$, which makes the
shortest domain shrink much faster than other domains. The flux  to the
right (resp. left) is proportional to $1/l_l$ (resp. $1/l_r$), which 
leads to the above ratio of $l_{0r}$ and $l_{0l}$.

\begin{figure}
\begin{center}
\epsfig{figure=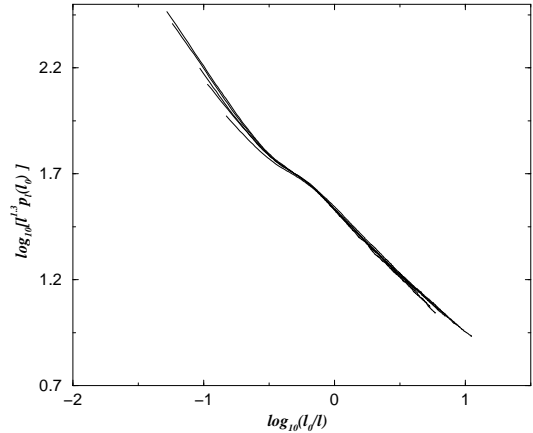,width=0.8\linewidth}
\end{center}
\caption{Scaling of the block persistence probability for the domain
aggregation model equivalent to the $1D$ zero temperature Cahn-Hilliard
dynamics. Simulations were carried out on a $10^6$ sites chain, for blocks
of size $101,131,161,191,211$. For this model $\lambda<d$ and the scaling 
function for block persistence has a power law divergence at small
$x$. Good scaling is obtained with $\theta'=1.3$, and we find $\theta=0.62$.}
\label{fig:cdet}
\end{figure}

Majumdar and Huse \cite{majumdar95a} have found numerically $\lambda=0.67$.
Therefore, this model is in the class $\lambda<d$ and we should observe 
a nontrivial $\theta'$. We have performed simulations of the domain model
on a chain of $10^6$ sites (20 samples), for blocks of size 101, 131, 161,
191 and 211. The scaling function presented in fig. \ref{fig:cdet} is 
qualitatively very different from the one in fig. \ref{fig:kawa_1D}.
 In agreement with the general discussion above and 
Eq. (\ref{eq:nontrivl}), we find
a cross-over between two exponents $\theta'>0$ and $\theta$. The best data 
collapse is obtained for $\theta'=1.3$, whereas we find $\theta=0.62$
(remark however that the large $x$ decay shows curvature).
It  may seem surprising that $\theta'>\theta$. However, since $\theta'$ is 
not directly related to the  the global persistence probability, there is
a priori no reason why  $\theta'$ should be less than $\theta$. 
 
\subsection{Two-dimensional Kawasaki dynamics}
\begin{figure}
\begin{center}
\epsfig{figure=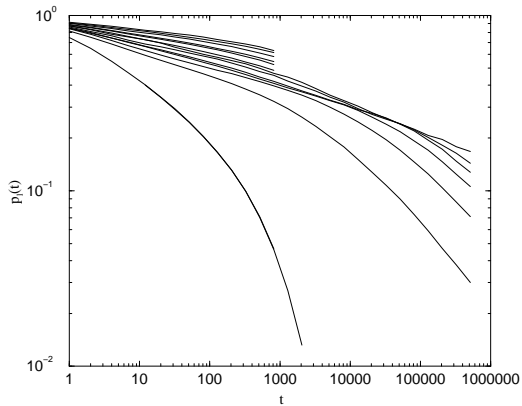,width=0.8\linewidth}
\caption{Results of simulations  of the Kawasaki spin-exchange dynamics on
a $700^2$ lattice, for blocks of size $l=$1, 3, 5, 7, 9, 11
and 13 (one sample, 500000 steps), 15, 21, 25, 35, 45, 55 (12 samples, 2000 steps).}
\label{fig:k2d}
\end{center}
\end{figure}

Effective zero-temperature domain models cannot be used successfully
 to avoid the freezing of the two-dimensional Kawasaki dynamics, because of
 the complicated geometry of the domains. Using block persistence
simulations can be performed at finite temperature. As mentioned above,
 simulations are difficult because the Kawasaki dynamics are very slow, and
do not reach the pure $t^{1/3}$ regime. Therefore, it is difficult to observe
block scaling and to extract the persistence exponent, and we have to be
 satisfied with qualitative results.
Figure \ref{fig:k2d} presents data obtained for a $1000^2$ systems with
a simulation time of $500000$ Monte-Carlo steps  for blocks of size 3, 5, 7,
 9, 11, and 15, and $2000$ steps for $l=$15, 21, 25, 35, 45 and 55.
 The cross-over in the behavior of $p_l(t)$
 corresponding to $L(t)\sim l$ is visible for small blocks.
 At large time, we observe a power
law decay with a persistence exponent $\theta\approx 0.5$. The actual value 
of $\theta$ is certainly bigger since the effective $z$ exponent increases 
with time and is still far from its asymptotic value $z=3$ ($1/z\approx 0.25$ 
at the end of the simulation). Still, we acknowledge that these data are not
 very conclusive.

It is not surprising to find a power law decay of the persistence 
probability (in the block scaling or $T\to 0$) limit, because of the slow
motion of interfaces, as for the nonconserved case.
What is less intuitive, is that
$\theta$ is much bigger for the conserved dynamics than for the nonconserved
dynamics.  However one can understand that  fast dynamics may lead to a
small $\theta$, if one realizes that a fast moving domain wall  
will be ineffective in decreasing $p(t)$ if it 
wipes several times regions of spins that have already flipped. 
Once again, we see that $\theta$ reflects very subtle effects.

At early times, when $l$ is increased, we do not seem to have a power law
 regime, but $\theta'=0$. 
 According to our discussion,
this would support the fact that $\lambda=d=2$ 
for  two-dimensional Kawasaki dynamics, but because $L(t)$ has strong
 corrections to scaling, it is not clear that correlations correctly scale
in the time regime observe, and one has to be careful. For  large blocks
and large time with $L(t)\sim t^{1/3}\ll l$, one might observe a power law.

\section{Conclusion}
In this article we have introduced the notion of  block persistence as 
a generalization of global and local persistence probabilities, and as a
way of giving a meaning to the persistence exponent $\theta$ {\it at finite
temperature}. Theoretical arguments as well as results of simulations
suggest that the persistence exponents and the whole scaling function of 
block persistence are temperature independent in the whole $T<T_c$ phase, which
is  conceptually speaking  very satisfactory. 
We have also shown that persistence
exponents  arise for conserved models as well, and that block persistence 
establishes a distinction between two classes of models, $\lambda=d$ and 
$\lambda<d$.

Finally, the important question may be: What do we learn from persistence ?
In fact, the justification for studying such crude models as the one
used in coarsening is universality, which states that most of the fine
details of the system are irrelevant for the study of the scaling regime.
Then the  stake is to identify universality classes, to understand 
the parameters  that determine them, and also to identify universal 
quantities. The theoretical and numerical study of persistence shows us,
because it   probes  temporal correlations very sensitively, that model 
universality   is not as wide as it may have been hoped {\it a
priori} from equilibrium-based intuition.
 While universality with respect to initial conditions, or
interactions range, seems to hold in most cases,  the present work suggests
 that  the continuous model $A$ and the Glauber Ising model in two 
dimensions are in
different universality classes (at least at zero temperature), even if they
have the same dimensionality, the same conservation law and both short-ranged 
interactions. Hence, the existence of the lattice seems to affect 
correlations even at large time. This could be related to the pertaining
of anisotropy at large time claimed by Rutenberg in a recent work 
\cite{rutenberg96a}.

The authors have benefitted from  interesting discussions with
S. Majumdar, B. Derrida, S. Cornell.

\bibliographystyle{prsty}
\bibliography{coarsening}

\end{multicols}
\end{document}